\documentclass[preprint,showpacs]{revtex4}
\usepackage{graphicx}% Include figure files
\usepackage{subfigure}
\begin{document}

\title[Dark energy, curvature, and cosmic coincidence]{Dark energy, curvature, and cosmic coincidence}

\author{Urbano Fran\c{c}a}

\affiliation{SISSA / ISAS, Astrophysics Sector, 
Via Beirut 4, 34014, Trieste, Italy\\
{\it and} \\
Instituto de F\'{i}sica Corpuscular  (CSIC-Universitat de Val\`{e}ncia),\\
Ed. Institutos de Investigaci\'{o}n, Apdo. 22085, E-46071, Valencia, Spain.}

\email{urbano.franca@ific.uv.es}

\begin{abstract}
The fact that the energy densities of dark energy and matter
are similar currently, known as the coincidence problem,
is one of the main unsolved problems of cosmology. We
present here a model in which a
spatial curvature of the universe can lead to a transition 
in the present epoch from a matter dominated universe to a scaling dark energy dominance 
in a very natural way. 
In particular, we show that if the exponential potential of the dark energy field
depends linearly on the spatial curvature density of a closed universe,
the observed values of some cosmological parameters can be obtained
assuming acceptable values for the present spatial curvature
of the universe, and without fine tuning in the only parameter
of the model. We also comment on possible variations
of this model, and realistic scenarios in which it could arise.
\end{abstract}
                        
%{\bf Keywords:} Dark energy theory, Cosmology of theories beyond the SM, Physics of the early Universe. }

\maketitle

%\vspace{.3cm}

\section{Introduction}

%{\it Introduction.} 
The cosmological scenario that has emerged in the last decade                        
indicates that we live in a universe which is almost spatially flat, as                        
indicated by the cosmic microwave background (CMB)                        
experiments \cite{wmap03,mactavish05,spergel06}. We have also learned                        
from observations \cite{turner03,tegmark04,seljak06} that its energy content is 
comprised by around 4$\%$ of baryons, and 26$\%$ of cold dark matter 
\cite{overduin04,bertone05},
both clumped in galaxies and clusters of galaxies. The remaining
70$\%$ of its energy is homogeneously distributed in the universe
and is causing its observed acceleration 
\cite{riess98,perl99,riess04,snls05}, a component
that was generically dubbed dark energy \cite{peebles03,pad03,copeland06}.

This recipe for the universe gives rise to 
what became known as the {\it cosmic 
coincidence
problem}: why are dark energy and dark matter energy densities of the 
same order of magnitude in the present epoch? This problem arises 
because 
dark energy must have a negative pressure to accelerate the universe, 
and 
cold dark matter (as well as baryons) has vanishing pressure. 
Therefore, the 
ratio of their energy densities $\rho$, for a constant equation of 
state $\omega_{\phi}$ of  dark energy,
must vary as
\begin {equation} 
\frac{\rho_M}{\rho_{\phi}} = \frac{\rho_{M0}}{\rho_{\phi 0}} 
(1+z)^{-3\omega_{\phi}} \approx \frac{(1+z)^3}{2} \ ,
\end {equation}
where $z$ is the redshift, 
$M$ denotes (barionic plus cold dark) matter, $\phi$ denotes the dark 
energy field, and the index 0 indicates the present value of a 
quantity. 

Several dark energy candidates have been proposed, from the most 
obvious, the cosmological constant \cite{weinberg89,carroll01}, to modifications 
of gravity \cite{cdg02,perrota00,carlo00,matarrese04}. 
Following the inflationary idea, "regular" scalar fields, the so-called
quintessence models, noncoupled 
\cite{wetterich88,ferreira97,ferreira98,ratra88, peebles88,frieman95,caldwell98} and coupled 
\cite{carroll97,casas92,farrar04,hoffman03,amendola00,amendola01,amendola03,
comelli03, prd04,biswas05,nojiri05} to dark 
matter content, were also proposed to be the cause of the acceleration, 
as well as fluids that do not obey the weak energy 
condition \cite{caldwell02,onemli02,caldwell03,carroll03,onemli04,feng05},  scalar fields with non-canonical 
kinetic terms \cite{picon00,picon01,chiba02,malquarti03,scherrer04} and  tachyonic 
fields \cite{pad02,sami02,bagla03,raul03,raul04}, just to cite a few examples.
These two last models have an advantage with
respect to others in what concerns the coincidence, 
in the sense that the 
equation of state of the field changes to a cosmological
constant-like as the background changes 
from radiation to  matter domination, but apparently 
they have a very small parameter space that can generate relevant
cosmological solutions \cite{raul03,raul04}.

In this sense, what models do, in general, is to fine tune 
the overall scale of the 
potential of dark energy (or the scale in which modifications of general 
relativity become
important) to be of order of the present critical density,
$ \rho_{c0} = 3 m_p^2 H_0^2 = 8.1 h^2 \times 10^{-47}$ GeV$^4$,   
where $H_0 = 100 h$ km s$^{-1}$ Mpc$^{-1}$ is the Hubble constant, 
$h= 0.72 \pm 0.08$ \cite{freedman01},
and $m_p = (8 \pi G)^{-1/2}= 1.221 \times 10^{18}$ GeV 
is the reduced Planck mass.

This fine tuning emerges even when one is using tracking or scaling 
\cite{copeland98,liddle98,zlatev99,steinhardt99}
properties of some potentials, like the exponential potential 
\cite{wetterich88,ferreira97,ferreira98,copeland98,jhep02,rubano04} 
we will focus on in this letter. In this sense, we have recently showed 
\cite{prd04} that coupling
dark energy with dark matter also does not solve the problem, 
since one has to adjust the value of the potential 
in very much the same way one does for the uncoupled quintessence.

Another general assumption when one is modelling dark
energy is that the universe is flat, that is,
curvature effects can be neglected. The Friedmann
equation in this case becomes
\begin {equation} \label{eq:friedmann} 
H^2 + \frac{k}{a^2} \approx H^2 = 
\left( \frac{\dot{a}}{a} \right)^2 =
\frac{1}{3 m_p^2} \left( \rho_M +
\rho_R + \rho_{\phi} \
\right) \ ,
\end {equation}
where $a(t)$ is the cosmic scale factor, dot 
indicates derivatives with respect to cosmic
time $t$, and $k$ is related to the curvature
via Robertson-Walker metric (+1, 0 or -1 for
a closed, flat or open universe, respectively). 
$H$ is the Hubble 
parameter, which present value is the Hubble 
constant, and R denotes the radiation
component. This equation
can then be rewritten as
%
%\begin {equation} \label{eq:omegas} 
$\Omega_M + \Omega_R + \Omega_{\phi} = 1 - \Omega_k 
\approx 1 $, 
%\ ,
%\end {equation}
%
where $\Omega_i = \rho_i / \rho_c = \rho_i / 3 m_p^2 H^2$
are the density parameters, and we have defined
$\rho_k \equiv - 3 m_p^2 k / a^2$. The energy densities
are given by the conservation of the energy-momentum
tensor for each component separately,
$ \dot{\rho}_i + 3 H \rho_i (1 + \omega_i) = 0 $,
where $\omega_M = 0$, $\omega_R = 1/3$ and $\omega_k = -1/3$.
The Lagrangian of the scalar field is the usual one,
$\mathcal{L} = \partial_{\mu} \phi \partial^{\mu} \phi/2 - 
V$,
and consequently, for a homogeneous perfect fluid, the equation
of state of dark energy is $\omega_{\phi} = p_{\phi} / \rho_{\phi}= 
(\dot{\phi}^2 /2 - V)/(\dot{\phi}^2 /2 + V)$ .

The flatness of the universe is one
of the main predictions of inflation
\cite{inflation}, since during an inflationary
period $\Omega_k$ vanishes quasi-exponentially.
However, recently some interest has been given to
the possibility that the spatial curvature 
of the universe be non-negligible \cite{linder05,polarski05,araujo05}, 
both for theoretical (see, for instance, 
\cite{barnard04,ellis04,freivogel05,adler05,lake05} and references therein)
and observational \cite{turner03,tegmark04,linder05,polarski05,white96,efstathiou03,uzan03} 
reasons. A positive
spatial curvature plus a cosmological constant
could, for instance, mimic a phantom regime 
\cite{linder05}. Another interesting 
point is that CMB experiments
\cite{wmap03,mactavish05,spergel06,seljak06}, even when
combined with different astronomical data
\cite{tegmark04,weller05,seljak06}, present a 
tendency for some small positive spatial
curvature of the universe, result that can
be checked soon using other observations
\cite{eisenstein98,blake03,matsubara04,eisenstein05,knox05,bernstein05,knox06,guzik06}. 

In this letter, we explore the possibility
that the universe has a global positive spatial curvature
in the context of a quintessence model. As
we will show,
if the potential is dependent on the curvature, 
there is no coincidence in the fact that
the transition from a matter dominated 
universe to a dark energy dominated universe
is happening currently, as well as the fact 
that the dark energy field behaves almost like
a cosmological constant. Instead, they are a
consequence of the fact the curvature is
becoming important in the present epoch. It should
be clear, however, we do not intend explaining 
why the curvature is becoming important right now. 
If observations really 
rule out flat models someday (what for sure 
is not the case today), this fact would be a different 
coincidence problem, the old flatness problem. We adopt here the
approach of checking what kind of solutions one
could obtain if observations indicate a non-flat
universe.

Once the dark energy attractor regime is
reached, after being triggered by the curvature,
the universe accelerates forever, and
the curvature again become negligible, in a 
kind of self-flattening process. In the following
section we describe the model, stressing 
its phenomenological nature. We then present an analytical
solution for the model, discuss its main
properties, and possible variants of it. Finally, in the conclusions
some realistic scenarios in which those models could arise
are briefly discussed.

\section{Dark energy and curvature}

%{\it Dark energy and curvature.} 
The models presented here are based on the assumptions 
that the universe presents a small positive curvature, and that the 
quintessential potential depends linearly on it, that is,
\begin{equation} \label{eq:pot}
V(\phi, \rho_k) = - \rho_k \ e^{\lambda \phi / m_p} 
= - \Omega_k \rho_c \ e^{\lambda \phi / m_p} \ ,
\end{equation}
where the negative sign comes from the fact $\Omega_k$
is negative for a closed universe, that is what we
consider in what follows. Since the "curvature density"
obeys the fluid-like equation \footnote{One should keep in mind that 
curvature is not a real fluid, and the
fact it obeys a fluid-like equation comes only from
the fact it scales as $a^{-2}$, not from the conservation of
a stress-energy tensor, like the other fluids.}, we can write
$V(\phi, u) = - \rho_{k0} \ e^{-2u} e^{\lambda \phi / m_p}$,
where $u= \ln (a/a_0) = - \ln (1+z)$, 
and $\rho_{k0} \equiv - 3 m_p^2 k / a_0^2$.

Based on the conditions, we will see that it is 
possible to explain
the current cosmological scenario with
natural values  for the only parameter of
the potential, $\lambda$, without fine tuning. 

It should be clear that such a dependence on
the curvature is phenomenological, 
although in the last section we point out
some contexts in which it may arise. In this section, however, 
we are interested in verifying what kind of
solutions one could obtain when such dependence 
is present. Notice also that modifying 
the exponential potential is not 
something new in the literature \cite{albrecht00,bento02},
although the dependence on the curvature is.

The scalar field equation can then be written as
\begin{equation} \label{eq:sf}
H^2 \phi'' + \frac{1}{2m_p^2} \left(
\rho_M + \frac{2}{3}\rho_R + 2 V +\frac{4}{3} \rho_k \right) \phi' =
\left(\frac{2}{\phi'} - \frac{\lambda}{m_p} \right) \ V \ ,
\end{equation}
where prime denotes derivatives with respect to $u$, and
the Hubble term, equation (\ref{eq:friedmann}), is given by
\begin{equation} \label{eq:hubble}
H^2  = \frac{\left( \rho_M + \rho_R + V + \rho_k \right) / 
3 m_p^2}{1 - {\phi'}^2/6 m_p^2} \ .
\end{equation}

\section{Results and discussion} \label{sec:results}

\begin{figure} 
%\begin{figure}%[ht]
%\vspace{0.5cm}
%\resizebox{!}{4cm}{\includegraphics{omegas.eps}}
%\resizebox{!}{4cm}{\includegraphics{eqest.eps}}
\includegraphics[scale=0.5]{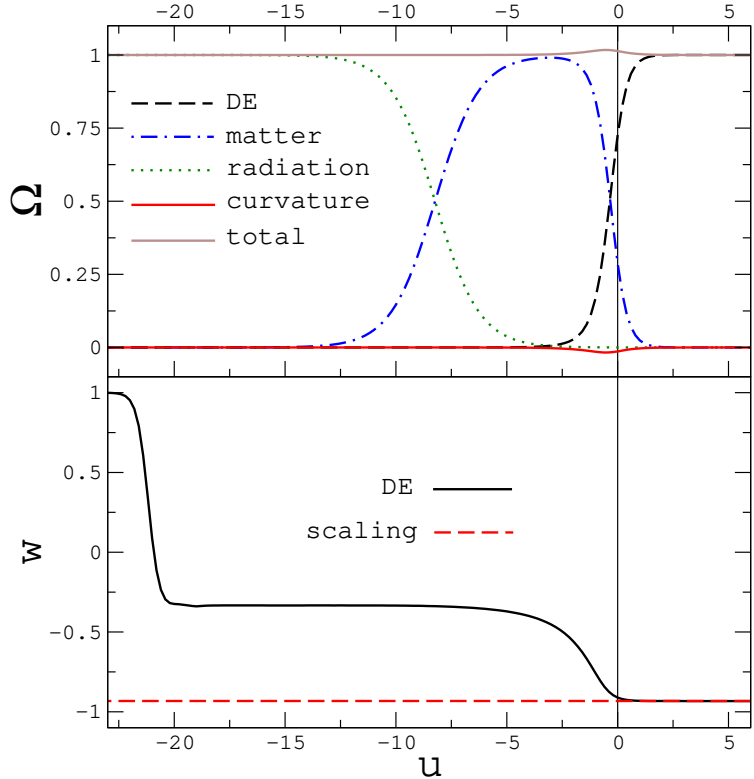}% Here is how to import EPS art
\caption{\label{fig:omegas} (Color online) {\it Top panel}: Density parameters 
of the 
components of the universe
as a function of $u=- \ln(1+z)$ for a typical solution (here, $\lambda=4$). 
Notice the transition to a dark energy dominated phase 
is happening by now, but on the contrary to other
models of quintessence, this is due to the fact the universe is
non-flat, and not to a fine tuning in the parameters.The upper solid 
curve indicated by {\tt total} corresponds to 
$\Omega_t = \Omega_R + \Omega_M + \Omega_{\phi} $. 
{\it Bottom panel}: Equation of state of the dark energy. Dashed curve
shows the scaling value, equation (\ref{eq:eos}).}
%and effective equation                        
%of state of the universe, equation (\ref{eq:effeos})
%.}
\end{figure}

%{\it Results and discussion.} 
The set of equations (\ref{eq:sf}) and (\ref{eq:hubble})
can then be solved numerically. It will
reach a scaling solution \cite{liddle98}, and therefore
the final solutions are almost independent of
the initial conditions. The results
for the density parameters of the components of
the universe for a typical solution are
shown in the top panel of figure \ref{fig:omegas}.  

%FIGURE 1
%%%%%%%%%%%%%%%%%%%%%%%%%%%%%%%%%%%%%%%%%%%%%%%%%%%%%%%%%%%%%%%%%%%%%%%%%%%%
%%%%%%%%%%%%%%%%%%%%%%%%%%%%%%%%%%%%%%%%%%%%%%%%%%%%%%%%%%%%%%%%%%%%%%%%%%%%%

Notice that the transition to the scaling regime is happening around
the present epoch, but differently of all models of
dark energy, there is no fine tuning in the overall scale
of the potential. Instead, the transition is triggered by
the fact that curvature is becoming important currently.
The system then reaches the accelerating scaling solution, as
can be seen in the bottom panel of the same figure, and when 
this happens the curvature again becomes negligible.

The scaling solution can be obtained from equations
(\ref{eq:sf}) and (\ref{eq:hubble}) when one realizes the field 
will dominate completely the energy density of the universe.
Since in a scaling solution we have $\phi'' = 0$, equation
(\ref{eq:sf}) becomes
%
%\begin{equation} \label{eq:hubbleshort}
$\frac{\phi'}{m_p}= \frac{2 m_p}{\phi'} - \lambda $,
%\end{equation}
%
which presents a solution,
\begin{equation} \label{eq:solution}
\frac{\phi'}{m_p}= \frac{\lambda}{2} 
\left( \sqrt{1 + \frac{8}{\lambda^2}} - 1 \right) \ .
\end{equation}
A more complete dynamical analysis will be presented
elsewhere. Here, we will focus
on this solution, that presents scaling properties, 
that is, the equation of state of the 
dark energy field is constant and different from
the one of background \cite{liddle98}. Its value
can be obtained using the fact that in the
scaling regime both kinetic and potential
terms of the scalar field scale in the same
way,
$ e^{-3(1 + \omega_{\phi})u} \propto
e^{-2u + \lambda \phi / m_p}$, and therefore, 
\begin{equation} \label{eq:eos}
\omega_{\phi} = - \frac{1}{3} - \frac{\lambda}{3} 
\frac{\phi'}{m_p} =  - \frac{1}{3} -
\frac{\lambda^2}{6} \left( \sqrt{1 + \frac{8}{\lambda^2}} - 1 \right) \ .
\end{equation}
The behavior of the equation of state for a typical
solution is presented in the bottom panel of figure 1,
for $\lambda=4$. For values of $\lambda \gg 2\sqrt{2}$, the dark energy
field practically mimics a cosmological constant. Contrary to what 
happens in the case of a regular exponential potential
\cite{wetterich88,ferreira97,ferreira98,copeland98,jhep02,rubano04}, 
where a large fine tuning on $\lambda$ is needed to obtain
$\omega_{\phi} \rightarrow -1$, here almost all values of
$\lambda$ generate acceptable values for the equation of state
of dark energy. In fact, expanding the square root of (\ref{eq:eos}) for small
$8/\lambda^2$ up to second order, one gets that
$\omega_{\phi} \approx -1 + 8/3\lambda^2$, from where one can see
it really approaches the cosmological constant value as
$\lambda$ increases.

%FIGURE ???
%%%%%%%%%%%%%%%%%%%%%%%%%%%%%%%%%%%%%%%%%%%%%%%%%%%%%%%%%%%%%%%%%%%%%%%%%%%%
%\begin{figure}%[ht]
%\vspace{0.5cm}
%\resizebox{!}{4cm}{\includegraphics{omegas.eps}}
%\resizebox{!}{4cm}{\includegraphics{eqest.eps}}
%\includegraphics[scale=0.9]{eos_lam.eps}% Here is how to import EPS art
%\caption{\label{fig:eoslam} \small Equation of state of 
%dark energy versus $\lambda$, given by equation (\ref{eq:eos}).}
%\end{figure}
%%%%%%%%%%%%%%%%%%%%%%%%%%%%%%%%%%%%%%%%%%%%%%%%%%%%%%%%%%%%%%%%%%%%%%%%%%%%%

However, the transition still is happening in the
present epoch, as we clearly can see from figure
\ref{fig:omegas}. Because of that, the scaling 
values have not yet been reached (although they almost have), 
and one needs 
to verify to which values of the curvature
reasonable cosmological parameters can be obtained.
In what follows, we looked for models with
$\Omega_{M0} = 0.3 \pm 0.1$ and $h = 0.72 \pm 0.08$
\cite{turner03,tegmark04,freedman01}. 

\begin{figure}  
%\vspace{0.5cm}
%\resizebox{!}{4cm}{\includegraphics{omegas.eps}}
%\resizebox{!}{4cm}{\includegraphics{eqest.eps}}
%\includegraphics[scale=0.56]{eos_lam_num.eps} \\% Here is how to import EPS 
\includegraphics[scale=0.5]{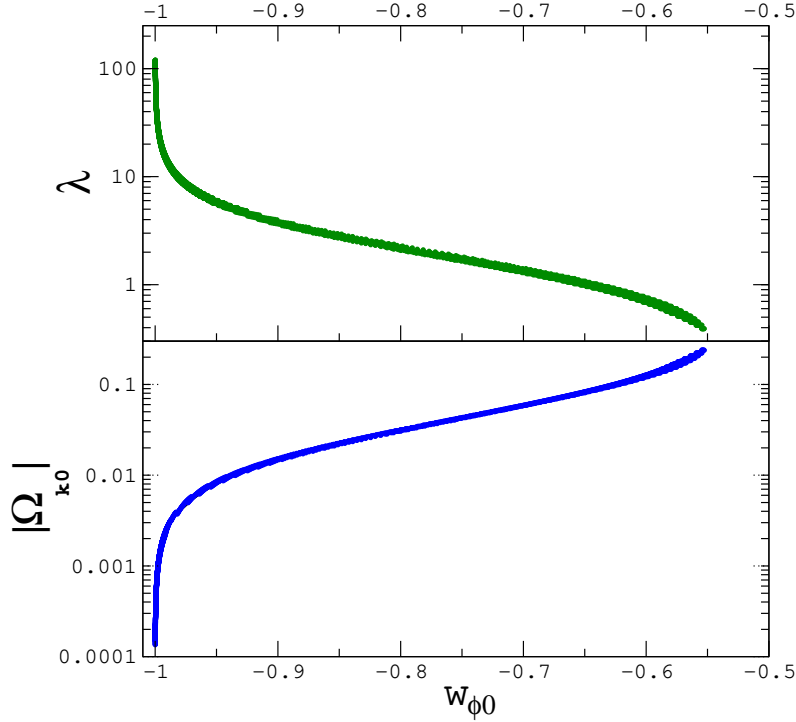}% Here is how to import EPS art
\caption{   \label{fig:eoslamnum} (Color online) {\it Top panel:} 
Current dark energy 
equation of state versus $\lambda$. 
{\it Bottom panel:}  Current dark energy 
equation of state versus the modulus of the present spatial 
curvature. Both figures are for models that satisfy the 
constraints
on $\Omega_{M0}$ and $h$ discussed in the text.}
\end{figure}

In order to do that, we have solved numerically
the system of equations varying 
$\Gamma_{k0} \equiv |\Omega_{k0}| h_{72}^2$ in the 
range $[10^{-6};0.2]$ 
%with stepsizes 
%$\Delta \Gamma_{k0} = 0.0015$ 
($h_{72} = h_n/0.72$, where $h_n$
is the Hubble constant obtained numerically), and
$\lambda$ in the range $[0.01,120]$, varying both 
stepsizes to get better resolution and faster
calculation.
%$\Delta \lambda = 0.03$.
The results are shown in figure \ref{fig:eoslamnum}.

%FIGURE 2
%%%%%%%%%%%%%%%%%%%%%%%%%%%%%%%%%%%%%%%%%%%%%%%%%%%%%%%%%%%%%%%%%%%%%%%%%%%%
%%%%%%%%%%%%%%%%%%%%%%%%%%%%%%%%%%%%%%%%%%%%%%%%%%%%%%%%%%%%%%%%%%%%%%%%%%%%%

Top panel shows the present value of the
equation of state of dark energy versus $\lambda$ for models
that satisfy the cited constraints. The values are not
exactly the ones given by equation (\ref{eq:eos}), 
since the transition has not been 
completed, but they are very close.  
Bottom panel presents the modulus of the spatial curvature today for the 
allowed models. A large curvature
would imply an equation of state incompatible
with observations. However, curious enough, for values of
the curvature within the current errors on its measured value,
the field presents an equation of state in agreement with
observations.

Before concluding, it should be pointed out 
that similar models might be considered, like for 
instance a modified Peebles-Ratra potential,
$V = - \rho_k (m_p / \phi)^{\alpha}$, or
another variations of the exponential potential like
$V = - \Omega_k m_p^4 e^{\lambda \phi / m_p}$, or
$V = - \rho_k (m_p / \phi)^{\alpha} e^{\lambda \phi^2 / m_p^2}$,
based on a variation of supergravity and 
supersymmetric models \cite{brax99,macorra05}.
Besides that, it might be clear that 
similar models can be constructed 
using $\Omega_k < 1$ (open models), although here
we have chosen to follow the indications given
by the CMB observations.

\section{Realistic models}

It is important to stress that the dependence
of the potential on the curvature is assumed 
here, and therefore it is crucial to verify whether it 
could be obtained in the context of a realistic 
particle physics model or in a modification of 
general relativity. In this sense, a 
particular direction which seems promising
is obtained in the context of models
with a scalar field non-minimally coupled to the
Ricci scalar \cite{nmc}. For a Friedmann-Robertson-Walker
universe, the Ricci scalar is given by 
$\mathcal{R}= -6 (\dot{H} + 2 H^2 + k/a^2)$, and 
it is conceivable that it does exist a Lagrangian which is a 
combination of non-minimal couplings
with the scalar invariants (like $F_1(\phi) \mathcal{R}$, 
$F_2(\phi) \mathcal{G}$, $F_3(\phi)/\mathcal{R}$, etc., where
$F_i(\phi)$ are functions of the field $\phi$ and 
$\mathcal{G}= R_{\mu \nu \lambda \rho}R^{\mu \nu \lambda \rho}
- 4 R_{\mu \nu}R^{\mu \nu} + \mathcal{R}^2$ is the so-called
Gauss-Bonnet term)  that can result on a dependence
of the potential with the curvature density, but not 
on $\dot{H}$ and $H^2$. If this is the case,
the dependence has the form
given in eq. (\ref{eq:pot}) \footnote{I acknowlegde 
Raul Abramo for discussions
on this point.}. 

Such kind of terms arise naturally in the extensions
of the standard model of particle physics, for instance in effective 
actions coming from superstring theory 
\cite{damour94,antoniadis94,hwang00,cartier01,hwang05,calcagni06,nojiri06}.
In string theory the field $\phi$ correspond to moduli fields (which 
generally appears in the form $\exp(\lambda \phi / m_p)$), and a 
combination of the several parameters appearing in such models 
\cite{hwang00,cartier01,hwang05,calcagni06}
also can potentially lead to the kind of cosmological solutions 
discussed here.

Another possible way is obtained in the so-called 
{\it macroscopic gravity approach} \cite{coley05}. The averaged
Einstein equations for a spatially {\it flat}, homogenous, and
isotropic universe have the form of normal Friedmann equations
plus a curvature term induced by the gravitational correlations 
terms. A comoving observer would therefore measure 
the universe to be spatially curved. In this way, 
a coupling of the scalar field with the 
correlation tensor could in principle lead to a 
coupling with a curvature-like term as the one studied here.

This is similar to what happens in the case of the {\it regional averaging}
\cite{buchert02,buchert03a}. In this case, the curvature-like term is 
generated by a volume effect and by curvature backreaction 
\cite{buchert03b,rasanen06}. Coupling the scalar field with the
new terms of the measured cosmological parameters might also result in a 
potential like the one given by equation (\ref{eq:pot}).
These and other possibilities are currently under investigation, 
and seem very promising to generate realistic models with the properties 
discussed here. 

One should note that, independently on the way the model
is generated, the general behavior of the models described here 
will be mantained. This can be seen from the fact 
that its main property (the fact that the spatial curvature triggers
the dark energy dominance) comes only from the fact
that the energy densities of matter, ``curvature'', and of the potential
energy of the dark energy field scales respectively as $a^{-3}, a^{-2}$, and
$a^{0}$. These scalings are independent of the gravity model, since two of them
come only from the conservation of the energy-momentum tensors, and the
other one only from the fact the curvature scales as $ a^{-2}$, which is 
a dimensional argument. Therefore,
the conclusions of the present work are expect to hold even for
more general theories and modifications of gravity, and should be 
considered as a possible alternative
to understand the coincidences which plague the present 
cosmological model.

\section{Conclusions} \label{sec:conc}

%{\it Conclusions.} 
We have presented a 
phenomenological model of dark energy 
in which its potential depends on the positive
spatial curvature of the universe, assumed
to be closed. The model presents a scaling
behavior that is triggered by the fact the
curvature is becoming non-negligible around the
present epoch, and, in this sense, there is
no coincidence in the fact $\Omega_{\phi 0} 
\approx \Omega_{M 0}$ \footnote{Although, as stated
before, one would need to explain 
why $\Omega_k$ is different
from zero only recently. Notice that this could be achieved, for instance,
using closed inflationary models: see, {\it e.g.}, 
\cite{ellis04,ellis91,linde95,ellis02,linde03,lasenby05} 
and references therein.}. 
Notice that it is a testable model, since the scenario 
would lose its appeal if observations indicate
$\Omega_{k0}$ is zero with enough accuracy.

In this sense, an important feature of the
model is that values of the
spatial curvature that give rise to reasonable 
cosmological parameters (like $\Omega_{M0}$,
$\omega_{\phi 0}$ and $h$) are within current
uncertainties in the observations of $\Omega_{k0}$.
Besides that, for all values
of $\lambda$ sufficiently high ($\lambda \gtrsim 2\sqrt{2}$)
the model behaves almost like a cosmological
constant, showing there is no fine tuning on this 
parameter. Further investigations
are needed to check if such a model would
still survive to tests from different observations, and 
more realistic models will allow one to infer
different features of such a scenario.

Of course it is important to be cautious and  
to keep in mind that, until now,
all the cosmological data are totally compatible
with a flat universe which energy
density is dominated by the cosmological constant. That
is probably the simplest scenario one can think 
to current cosmology, ignoring the coincidence.
Whether alternatives like the one described here 
are cosmologically reasonable is something that depends
on the forthcoming observations, especially the ones able 
to probe a possible deviation from
flatness on the spatial curvature of universe. 

\section*{Acknowledgments}
I would like to acknowledge Raul Abramo, 
Carlo Baccigalupi, Ana Mosquera, Massimo Pietroni, Rogerio
Rosenfeld, and Thomas Sotiriou for several discussions and 
comments on the manuscript.

\end{document}